\documentclass[12pt]{article}
\usepackage{amssymb}

\begin{document}

\begin{center}
\bigskip

\bigskip

{\large \textbf{Gravitational Noise with Polarization Variables, the Nature
of Entanglement States and Metrics Fluctuational Interpretation of Qauntum
Mechanics}}

\medskip T.~F.~Kamalov

Physics Department, Moscow State Opened University,

ul. P. Korchagina, 22, Moscow 107996, Russia

Tel./fax 7-095-2821444

\begin{tabular}{ll}
E-mail: & okamalov@chat.ru \\ 
& ykamalov@rambler.ru
\end{tabular}
\end{center}

{\small
Discussed in the study are gravitational noise and the nature of
entanglement states. Their role in forming of entangled states. Nonlocal
nature of entangled states can be brought about by Polarization Variables.
Polarization Variables consists of sum a background of random classical
gravitational fields and waves, random electromagnets fields and so on known
or unknown, distributed average isotropically in the space. This background
is capable of correlating phases the oscillations of microobjects. From this
follow, that entanglement polarizations states is the functions of
Polarization Variables in Vacuum.
}

\bigskip

Pacs 03.65*

Key words: Metrics Fluctuational Interpratation,Gravity Background, Gravity
Noise, Polarization Variables, Entanglement State.
\bigskip

\everymath{\displaystyle}
\bigskipamount=.1\bigskipamount

The subject of present study is polarization in vacuum and influence of
vacuum polarization variables onto origination of entangled states. The same
reasons could be employed to explain nonlocal character of entangled states.

In the physical vacuum, there always exist a gravitational background, that
is, the background of stochastic gravitational fields and waves, as
gravitational fields cannot be shielded. Therefore, the energy and the
action function for classic vacuum are not zeros. If there exist a
quasi-classic approach for transition from the quantum particle to a classic
one and vice versa, the one can impose the requirement of equality of action
function for classic and quantum vacuum, as these constitute different
descriptions of the same physical object.

Physical vacuum is understood as a volume from which everything possible has
been removed. With this, depending on the type of this physical object
description, different definitions are adopted for the latter. In case of
classic description via classic physics, one can state the impossibility of
removal therefrom classic gravitational fields and waves. Hence, the classic
vacuum can be considered possessing the nonzero energy and the action
function. On the other hand, when regarding the quantum vacuum, its energy
and action function are considered to be nonzero even without accounting for
gravitational interactions. In this case, the influence of the gravitational
background is usually neglected due to its insignificantly low value. It
should be noted that by present, the gravitational background has not been
studied either theoretically or experimentally.

Theoretical investigation of the vacuum accounting for gravitational fields
has been performed in the studies of Academician Andrey D.Sakharov [1, 2].
In his first work of 1967, ''Vacuum quantum fluctuations in the curved space
and gravitation theory'' it has been stated that ''it is assumed in the
modern quantum field theory that the energy-momentum tensor of vacuum
quantum fluctuations equal zero , and the respective action $S(0)$ is
actually zero''. He has further shown that accounting for gravitational
field in the vacuum, taking into consideration definition of space-time
action dependence form curvature in the gravitational theory of A.Einstein
(with invariants of Ricci tensor $R$ and metric tensor $g$), the action
function shall take on the form

\begin{center}
$S(R)=-\frac{1}{16\pi G}\int (dx)\sqrt{-g}R$.
\end{center}

Resultant action of all these gravitational fields with number $j$ form the
functional

\begin{center}
$S_{0}(\psi )=\sum_{j=1}^{\infty }S_{j}$,
\end{center}

\noindent
being the external field $\psi (x)$ given by the metric tensor $g_{ik}$ of
the gravitational field. This brings about the phenomenon he has called
vacuum polarization. He believed this effect could be manifested for quantum
objects.

Regretfully, these ideas could not be developed at that time due to
experimental base having not reached the level adequate for verification of
such effects.

In the present study, we suggest to employ the idea of polarization
variables in vacuum for explanation of the nature of entangled quantum
states and of the nonlocal nature of Bell's inequalities. To seam quantum
and classic vacuums, let us require equality of action functions for quantum
and classic vacuums:

\begin{center}
$S_{0}=\sum_{j=1}^{\infty }S_{j}=-\frac{1}{16\pi G}\sum_{j=1}^{\infty }\int
(dx)\sqrt{-g_{j}}R_{j}=\hbar $,
\end{center}

\noindent
$S_{0}$ being the resultant action function for gravitational backgrounds
with gravitational wave or field number $j$.

When considering Bell's inequalities and entangled quantum states, what is
measured is correlation of polarization states of quantum particle entangled
states. At the same time, it is not quite clear where the strange effect of
interdependence of quantum particle entangled states originate from, as
these particles are assumed to be non-interacting. The answer to this
question follows from Academician Sakharov's vacuum polarization.
Gravitational fields through influencing non-interacting microobjects is
binding their polarization states [3, 4].

Let us consider two classic particles in a field of random gravitational
fields or waves. The General Theory of Relativity gives the length element
in 4-dimensional Riemann space as

\begin{center}
$\Delta \ell ^{2}=g_{ik}\Delta x^{i}\Delta x^{k}$,
\end{center}

\noindent
the metric in the linear approach is

\begin{center}
$g_{ik}=\eta _{ik}+h_{ik}$,
\end{center}

\noindent
being $\eta _{ik}$ Minkowsky metric, constituting the unity diagonal matrix.
Hereinafter, the indices $\mu ,\nu ,\gamma $ acquire values 0, 1, 2, 3.
Indices encountered twice imply summation thereupon. Let us select harmonic
coordinates (the condition of harmonicity of coordinates mean selection of
concomitant frame $\frac{\partial h_{n}^{m}}{\partial x^{m}}=\frac{1}{2}%
\frac{\partial h_{m}^{m}}{\partial x^{n}}$) and let us take into
consideration that $h_{\mu \nu }$ satisfies the gravitational field equations

\begin{center}
\bigskip $\square h_{mn}=-16\pi GS_{mn}$,
\end{center}

\noindent
which follow from the General Theory of Relativity; here $S_{mn}$ is
energy-momentum tensor of gravitational field sources with d'Alemberian $%
\square $ and gravity constant $G$. Then, the solution shall acquire the form

\begin{center}
$h_{\mu \nu }=e_{\mu \nu }\exp (ik_{\gamma }x^{\gamma })+e_{\mu \nu }^{\ast
}\exp (ik_{\gamma }x^{\gamma })$,
\end{center}

\noindent
where the value $h_{\mu \nu }$\ is called metric perturbation, $e_{\mu \nu }$%
\ polarization, and $k_{\gamma }$\ is 4-dimensional wave vector. We shall
assume that , which constitute metric perturbation $h_{\mu \nu }$, are
distributed in space with an unknown distribution function $\rho =\rho
(h_{\mu \nu })$.

Relative oscillations $\ell $ of two particles in classic gravitational
fields are described in the General Theory of Relativity by deviation
equations

\begin{center}
$\frac{D^{2}}{D\tau ^{2}}\ell ^{i}=R_{kmn}^{i}\ell ^{m}\frac{dx^{k}}{d\tau }%
\frac{dx^{n}}{d\tau }$,
\end{center}

\noindent
being $R_{kmn}^{i}$ gravitational field Riemann's tensor.

Specifically, the deviation equations give the equations for two particles
oscillations

\begin{center}
$\stackrel{..}{\ell }^{1}+c^{2}R_{010}^{1}\ell ^{1}=0,\quad \omega =c\sqrt{%
R_{010}^{1}}$.
\end{center}

The solution of this equation has the form

\begin{center}
$\ell ^{1}=\ell _{0}\exp (k_{a}x^{a}+i\omega t)$,
\end{center}

\noindent
being $a=1,2,3$. Each resultant gravitational field or wave with index $j$
and Riemann's tensor $R_{kmn}^{i}(j)$ shall be corresponding to the value $%
\ell ^{i}(j)$ with stochastically modulated phase $\Phi (j)=\omega (j)t$.

We assume to be dealing with a stochastic gravitational background, so we
are to define the probability in 4-dimensional Riemann space. This
definition shall satisfy the following requirements.

We shall call the stochastic curved space the 4-dimensional Riemann space
with probabilities defined therein. With this, we shall require the
following:

1. If the interval $\Delta \ell $ in the Riemann space equals zero, then the
probability of finding the particle in this interval equals unity.

2. If the interval $\Delta \ell $ in the Riemann space equals infinity, then
the probability of finding the particle in this interval equals zero.

3. The probability to find the particle in the intervals $\Delta \ell
(x_{2}^{i},x_{1}^{i})+\Delta \ell (x_{3}^{i},x_{2}^{i})\geq \Delta \ell
(x_{3}^{i},x_{1}^{i})$, equals $P_{21}+P_{32}\leq P$.

Hereinafter, we refer to the regions of 4-dimensional Riemann space, which
are bount with each other with single type intervals, that is, only with
ether time-like intervals or space-like intervals.

It should be noted that these requirements are met by intervals with the
following type of distribution: interval $\Delta \ell $ is corresponding
with the probability interval

\begin{center}
$\Delta P=\frac{1}{\sigma \sqrt{2\pi }}\exp (-\frac{\Delta \ell ^{2}}{%
2\sigma ^{2}})$.
\end{center}

Similarly, for the probability interval of finding a particle with velocity $%
u^{i}$\ we have:

\begin{center}
\bigskip $\Delta P=\frac{1}{\sigma \sqrt{2\pi }}\exp (-\frac{S}{S_{0}})$,
\end{center}

\noindent
or

\begin{center}
$\Delta P=\frac{1}{\sigma \sqrt{2\pi }}\exp (-\frac{W}{m\sigma ^{2}})$ ,
\end{center}

being $W$ the energy of oscillation of the micro-particle in question
acquired in the gravitational background; for a free particle with the RMS
velocity deviation $(\Delta u^{i})^{2}$\ this energy equals action function $%
S$. Assuming the dispersion to be characteristic for the gravitational
background and formally equating the dispersion with gravitational
background energy, we obtain

\begin{center}
$\Delta P=-\frac{1}{\sigma \sqrt{2\pi }}\exp (-\frac{S}{S_{0}})$.
\end{center}

Then, introducing notations $a^{2}=\frac{1}{\sigma \sqrt{2\pi }}$, $S_{0}=%
\frac{m\sigma ^{2}}{2}$ and taking into account that the probability is
equal to squared probability amplitude, the probability amplitude,
characterizing the action acquired by any micro-particle of mass $m$\ in the
gravitational background, will equal

\begin{center}
$\psi =a\exp (i\frac{S}{S_{0}})$.
\end{center}

The probability amplitude is a vector in the complex space. Therefore, it
can be decomposed in vector basis $e^{m}$, so that

\begin{center}
$\psi =e^{m}\psi _{m}$.
\end{center}

The scalar product of two vectors $A^{m}$ and $B^{n}$in the Riemann space is
defined as $g_{mn}A^{m}B^{n}$ , hence we can normalize this wave function as
follows:

\begin{center}
$\int g_{mn}\psi _{m}\psi _{n}dx=1$.
\end{center}

If we pass to the 3-dimensional space, having added the classical 
Jacobi--Hamilton equation

\begin{center}
\bigskip $\frac{\partial S}{dt}+\frac{1}{2m}(\nabla S)^{2}+U=0$
\end{center}

\noindent
and continuity equation for probability density $\psi ^{2}=a^{2}$,

\begin{center}
$\frac{\partial a^{2}}{\partial t}+div(a^{2}\frac{\nabla S}{m})=0$,
\end{center}

\noindent
we shall get the Schroedinger equation for this wave function

\begin{center}
$i2S_{0}\frac{\partial \psi }{\partial t}=-\frac{4S_{0}^{2}}{2m}%
\bigtriangleup \psi +U(x,y,z)\psi $.
\end{center}

Next, let us pass to Bell's inequalities [5].

We shall consider the physical model with the gravitational noise [i.e. with
the background of gravitational fields and waves]. This means that we assume
existence of fluctuations in gravitational waves and fields expressed
mathematically by metric fluctuations.

Considering quantum micro-objects in the curved space, we shall take into
consideration the fact that the scalar product of two 4-vectors $A^{i}$ and $%
B^{k}$ equals $g_{ik}A^{i}B^{k}$, where for weak gravitational fields one
can use the value $h_{\mu \nu }$, which is the solution of Einstein's
equations for the case of weak gravitational field in harmonic coordinates.

Correlation factor $M$\ of random variables $\lambda ^{i}$\ projections onto
directions $A^{k}$\ and $B^{n}$\ defined by polarizers (all these being
unity vectors) is

\begin{center}
$\left| M\right| =\left| \left\langle \lambda ^{i}A^{k}g_{ik}\lambda
^{m}B^{n}g_{mn}\right\rangle \right| =\left| \left\langle
g_{ik}g_{mn}\lambda ^{i}\lambda ^{m}A^{k}A^{n}\exp i\theta \right\rangle
\right| =$

$=\left| \left\langle g\exp i\theta \right\rangle \right| =\left| \exp
i\theta \right| $
\end{center}

\noindent
averaging gives $\left\langle g\right\rangle =1$; in the introduced
notation, $g=g_{ik}g_{mn}\lambda ^{i}\lambda ^{m}A^{k}A^{n}$. Here, all
vectors are considered to be unity ones, while averaging of the metric in
the weak field approach gives unity. is the angle $\theta $\ between
polarizers, hence $B^{n}=A^{n}e^{i\theta }$, indices $i,k,m,n$\ acquiring
values 0, 1, 2, 3.Finally, the real part of the correlation factor is

\begin{center}
$\left| M_{AB}\right| =\left| \cos \theta \right| $.
\end{center}

Then, the Bell's observable $S$ in Rieman's space for $\theta =\frac{\pi }{4}
$, we obtain the maximum value of

\begin{center}
$\left| <S>\right| =\left| \frac{1}{2}[\langle M_{AB}\rangle +\left\langle
M_{A^{\prime }B}\right\rangle +\left\langle M_{AB^{\prime }}\right\rangle
-\left\langle M_{A^{\prime }B^{\prime }}\right\rangle ]\right| =$

$=\left| \frac{1}{2}[\cos (-\frac{\pi }{4})+\cos (\frac{\pi }{4})+\cos (%
\frac{\pi }{4})-\cos (\frac{3\pi }{4})]\right| =\left| \sqrt{2}\right| $,
\end{center}

\noindent
which agrees fairly with the experimental data. The Bell in equality in
Rieman's space shall take on the form ${\left| \left\langle S\right\rangle
\right| \leq \sqrt{2}}$.

Presently, there is very little reliable information about gravitational
background. The latter is not investigated either theoretically or
experimentally. However, one can expect that the due account of it is
capable of providing answers for a number of questions in the modern quantum
theory. The role of gravitational polarization variable onto formation of
entangled states testifies to the possibility of their obtaining through
introduction of these concepts into Bell's inequalities.

I'm thanks to Doctor Yuriy Svirko from Tokyo University, Department of
Applied Physics and to Professor Yuriy Rybakov from People's
Friendship University of Russia, department of Theoretical Physics for help
and assistant.

\end{document}